\documentclass[pdflatex,sn-nature]{sn-jnl}

\usepackage{natbib}
\usepackage[T1]{fontenc}
\usepackage{graphicx,epsfig} 
\usepackage{multirow}%
\usepackage{amsmath,amssymb,amsfonts}%
\usepackage{amsthm}%
\usepackage{mathrsfs}%
\usepackage[title]{appendix}%
\usepackage{xcolor}%
\usepackage{textcomp}%
\usepackage{manyfoot}%
\usepackage{lineno} 
\usepackage{booktabs}%
\usepackage[hyphens]{xurl}
\usepackage{pdfpages}

\usepackage{xcolor}
\newcommand{\rev}[1]{\textcolor{black}{#1}}

\begin{document}

\title[Article Title]{Geometric representations of brain networks can predict the surgery outcome in temporal lobe epilepsy}

\author*[1]{\fnm{Martin} \sur{Guillemaud}}\email{martin.guillemaud@gmail.com}

\author[1]{\fnm{Alice} \sur{Longhena}}

\author[1,2]{\fnm{Louis} \sur{Cousyn}}

\author[1,2]{\fnm{Valerio} \sur{Frazzini}}

\author[1,3]{\fnm{Bertrand} \sur{Mathon}}

\author[1,2]{\fnm{Vincent} \sur{Navarro}}

\author[4]{\fnm{Mario} \sur{Chavez}}

\affil*[1]{\orgname{Paris Brain Institute (ICM), CNRS, Inserm, Sorbonne University, Inria-Paris. Pitié Salpêtrière Hospital}, \orgaddress{\city{Paris}, \country{France}}}

\affil[2]{ \orgname{AP-HP, Department of Neurology, Epilepsy Unit, Center of Reference for Rare Epilepsies, ERN EPICARE, Pitié Salpêtrière Hospital}, \orgaddress{\city{Paris}, \country{France}}}

\affil[3]{ \orgname{AP-HP, Department of Neurosurgery, Pitié Salpêtrière Hospital}, \orgaddress{\city{Paris}, \country{France}}}

\affil[4]{\orgname{CNRS, Pitié Salpêtrière Hospital}, \orgaddress{\city{Paris}, \country{France}}}

\abstract{Epilepsy surgery, particularly for temporal lobe epilepsy (TLE), remains a vital treatment option for patients with drug-resistant seizures. However, accurately predicting surgical outcomes remains a significant challenge. This study introduces a novel biomarker derived from brain connectivity, analyzed using non-Euclidean network geometry, to predict the surgery outcome in TLE. Using structural and diffusion magnetic resonance imaging (MRI) data from 51 patients, we examined differences in structural connectivity networks associated to surgical outcomes. Our approach uniquely utilized hyperbolic embeddings of pre- and post-surgery brain networks, successfully distinguishing patients with favorable outcomes from those with poor outcomes. Notably, the method identified regions in the contralateral hemisphere relative to the epileptogenic zone, whose connectivity patterns emerged as a potential biomarker for favorable surgical outcomes. The prediction model achieves an area under the curve (AUC) of 0.87 and a balanced accuracy of 0.81. These results underscore the predictive capability of our model and its effectiveness in individual outcome forecasting based on structural network changes. Our findings highlight the value of non-Euclidean representation of brain networks in gaining deeper insights into connectivity alterations in epilepsy, and advancing personalized prediction of surgical outcomes in TLE.}

\keywords{Epilepsy, brain networks, hyperbolic geometry, surgery outcome prediction}

\maketitle

\section{Introduction}\label{Sec:intro}
Epilepsy is one of the most prevalent neurological disorders, affecting approximately 1\% of the global population ~\cite{chan2003mechanisms}. Temporal lobe epilepsy (TLE) is the most common form of drug-resistant focal epilepsy, often leading to a significant reduction in quality of life due to recurrent and unpredictable seizures ~\cite{chan2003mechanisms}. For drug-resistant epilepsies, which comprise about 30\% of cases, resective epilepsy surgery has become a widely accepted therapeutic option aiming at removing the epileptogenic regions~\cite{Wiebe2001, de2011long, Jobst2015}. However, approximately 30\% of these patients continue to experience persistent seizures after surgery ~\cite{de2011long}. One potential explanation is that seizures may originate from abnormal brain regions that were not resected. Emerging evidence suggests that in TLE, structural abnormalities extend beyond the epileptogenic zone, forming a broader network involved in seizure generation~\cite{Bernhardt2013, besson2014structural}. These studies indicate that a more comprehensive understanding of brain connectivity changes associated with favorable surgical outcomes could enhance surgical planning and postoperative care~\cite{Hsieh2023, Spencer2008}.

Although some scoring systems correlate with surgical outcomes ~\cite{Radhakrishnan1996, Tonini2004, Gracia2015}, existing models remain insufficient to reliably guide clinicians in predicting surgical success, leaving a gap in personalized treatment strategies~\cite{uijl2008prognosis, Gracia2019}. Over recent decades, clinical data have been combined with markers from preoperative magnetic resonance imaging (MRI), scalp electroencephalography (EEG), intracranial EEG (iEEG), or magnetoencephalography (MEG) data, to develop predictive tools~\cite{Armananzas2013, Memarian2015, Gleichgerrcht2018, aydin2020magnetoencephalography, Sinha2023}. Features from interictal pathological iEEG activity, such as spikes and high-frequency events in the epileptogenic zone, have been used as biomarkers of surgical success~\cite{haegelen2013high, thomas2023subpopulation}. Connectivity patterns of these events, however, provide more accurate outcome predictions~\cite{gonzalez2019high, lin2024high}. For example, preoperative iEEG analyses show that resecting weakly homogeneous networks in TLE often leads to poor outcomes~\cite{antony2013functional}. Non-invasive studies have demonstrated the predictive utility of spectral power and coherence features from presurgical scalp EEG~\cite{varatharajah2022quantitative, sheikh2024machine}, and MEG-derived cortical networks, with localized epileptic regions correlating with seizure-free outcomes~\cite{aydin2020magnetoencephalography}. Structural abnormalities in MRI, including the morphology of temporal structures, also predict surgical success when compared to normative data~\cite{feis2013prediction, bernhardt2015magnetic, morita2021incorporation}. Functional connectivity from presurgical fMRI has been associated with postoperative seizure freedom~\cite{lariviere2020functional}, with poor outcomes linked to regional network segregation~\cite{DeSalvo2020}. 

Interestingly, recent evidence shows that integrating connectivity information from fMRI and diffusion MRI data could serve as a reliable biomarker for predicting surgical outcomes~\cite{morgan2022presurgical}. Several studies have linked presurgical white matter properties to surgical outcomes, including tract density~\cite{alizadeh2019hemispheric} and diffusion abnormalities in white matter bundles from DTI data~\cite{keller2017preoperative}. Patient-specific white matter features from dMRI data have been shown to be reliable biomarkers of postoperative seizure outcomes~\cite{Bonilha2015}. Structural networks derived from DTI provide accurate, individualized outcome predictions~\cite{Gleichgerrcht2018, gleichgerrcht2020temporal, Sinha2021, Johnson2022}. Surgical outcomes depend not only on presurgical networks but also on how resection affects brain connectivity~\cite{taylor2018impact, da2020network, Sinha2021}. Predictive modeling of multivariate iEEG data has enabled seizure propensity predictions by simulating channel resection~\cite{steimer2017predictive, muller2018evaluating}. Network-based in-silico simulations assess the ``ictogenicity'' of brain areas and predict outcomes of virtual resections~\cite{hutchings2015predicting, goodfellow2016estimation, sinha2017predicting, kini2019virtual}.  Our study leverages MRI and dMRI data to analyze pre- and postoperative structural brain connectivity changes, aiming to identify novel biomarkers for more precise surgical outcome predictions.

In recent years, brain connectivity networks have emerged as a powerful framework for studying a range of neurological diseases, including neurodegenerative disorders~\cite{seeley2009neurodegenerative, Perovnik2023}, schizophrenia~\cite{van2014brain}, and epilepsy~\cite{chiang2014graph, gleichgerrcht2015connectomics}. Representing the brain as a network of nodes (e.g., brain regions, sensors, voxels) and edges (functional or anatomical connections) offers a comprehensive view of brain architecture, surpassing traditional region-based approaches~\cite{reijneveld2007application, stam2014modern, fornito2016fundamentals}. Rather than identifying a single cortical area responsible for seizures, network-based approaches have highlighted the critical role of widespread altered connectivity beyond the epileptogenic zone~\cite{Richardson2012, chiang2014graph, besson2014structural}.

Brain connectivity networks are typically represented as adjacency matrices or edge lists, but their high dimensionality complicates statistical analysis (e.g., node classification, clustering, and link prediction)~\cite{fornito2016fundamentals}. To simplify analysis, networks are generally projected into low-dimensional vector spaces while preserving their structural properties. Although Euclidean embeddings are commonly used, they often require high dimensions and fail to capture key features of large-scale brain networks like hierarchical structure~\cite{Nickel2017, Sala2018, Whi2022}.

Hyperbolic graph embedding has gained attention for its effectiveness in representing complex networks. Unlike Euclidean space, where distances grow linearly, hyperbolic space features exponentially expanding distances, making it ideal for capturing hierarchical and scale-free structures common in real-world networks~\cite{Krioukov2010}. This allows for lower distortion embeddings that preserve both local and global connectivity structures more effectively than Euclidean projections~\cite{Papadopoulos2012, Boguna2008}.

Hyperbolic embeddings offer key advantages for studying brain connectivity. Brain networks exhibit a small-world and hierarchical structure, with local clusters of tightly connected regions and long-range connections integrating functional modules~\cite{fornito2016fundamentals}. Hyperbolic space naturally captures these properties in a lower-dimensional space~\cite{Allard2020}.  Recent studies have shown that hyperbolic embeddings are effective for exploring brain network disruptions in neurological conditions. For example, hyperbolic embedding has been used to investigate brain network alterations linked to cognitive decline in Alzheimer's disease ~\cite{Baker2024, Longhena2024bis}, and autism spectrum disorder~\cite{Whi2022}.

In epilepsy research, hyperbolic mapping of brain networks has shown potential for localizing connectivity disruptions caused by surgery~\cite{Longhena2024}, and identifying brain states at high risk of seizures~\cite{Guillemaud2024}. In this study, we used hyperbolic graph embedding to analyze pre- and post-surgical brain networks in 51 patients who underwent anterior temporal lobe resection (ATLR) surgery~\cite{Sinha2021}. Connectivity networks were constructed from diffusion and structural MRI data before and after surgery. Embedding these networks into hyperbolic space enabled a direct comparison of connectivity changes linked to surgical outcomes, allowing us to assess whether pre- and post-surgical network differences could serve as biomarkers for favorable outcomes.

Analysis of  embedded networks allowed us to evaluate how surgical resection affects brain connectivity by identifying both short- and long-range effects around the surgical region. This approach highlighted specific brain regions contributing to surgical outcome differentiation, providing insights into the anatomical and network-level changes linked to surgical success. In addition to comparing network structures, hyperbolic embeddings were used to build a predictive model for surgical outcomes. Our approach yielded a good prediction performance (area under the curve AUC$= 0.87$). This performance was significantly enhanced (AUC$=0.90$) for patients who underwent a left hemisphere surgery. Our results demonstrate that hyperbolic geometry offers a novel framework for analyzing brain network changes in TLE surgery, with potential to improve understanding of surgical effects and patient-specific outcome prediction.

\section{Results}\label{Sec:results}
\subsection{Comparison of pre- and post-surgery brain networks}
\begin{figure}
    \begin{center}
    \includegraphics[width=1\textwidth]{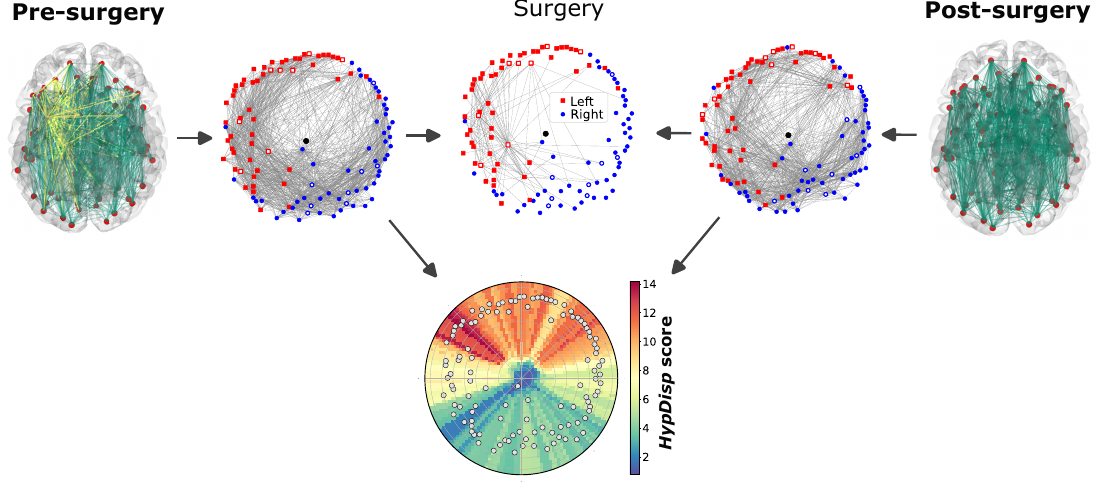}
    \caption{Embedding and comparison of pre- and post-operative networks in the hyperbolic disk. Pre- and post-surgery brain networks are embedded and aligned in the disk. The $\mathrm{HypDisp}$ score is calculated for each node and interpolated across the disk for each patient. Red squares represent left hemisphere nodes, blue dots represent right hemisphere nodes, and the black dot marks the origin $(0,0)$ of the disk. White-faced nodes correspond to the temporal lobes (left or right, based on edge color). The surgery network represents the connections that were removed during the surgery.}
    \label{data_Embedding}
    \end{center}
    \end{figure}

The impact of surgery on brain connectivity was quantified by comparing the displacement of a node in the embedding space caused by a network perturbation. To quantify this change, we  estimated the $\mathrm{HypDisp}$ score of each node from the pre- and post-surgery networks embedded in the hyperbolic disk~\cite{Longhena2024}. This displacement score is based on the idea that a local perturbation in a node’s connectivity will alter its embedding coordinates relative to those of the original graph. Fig.~\ref{data_Embedding} illustrates the procedure for one patient. Interestingly, the embedding separates nodes corresponding to the left and right hemispheres in an unsupervised manner.

\begin{figure}
    \begin{center}
    \includegraphics[width=1\textwidth]{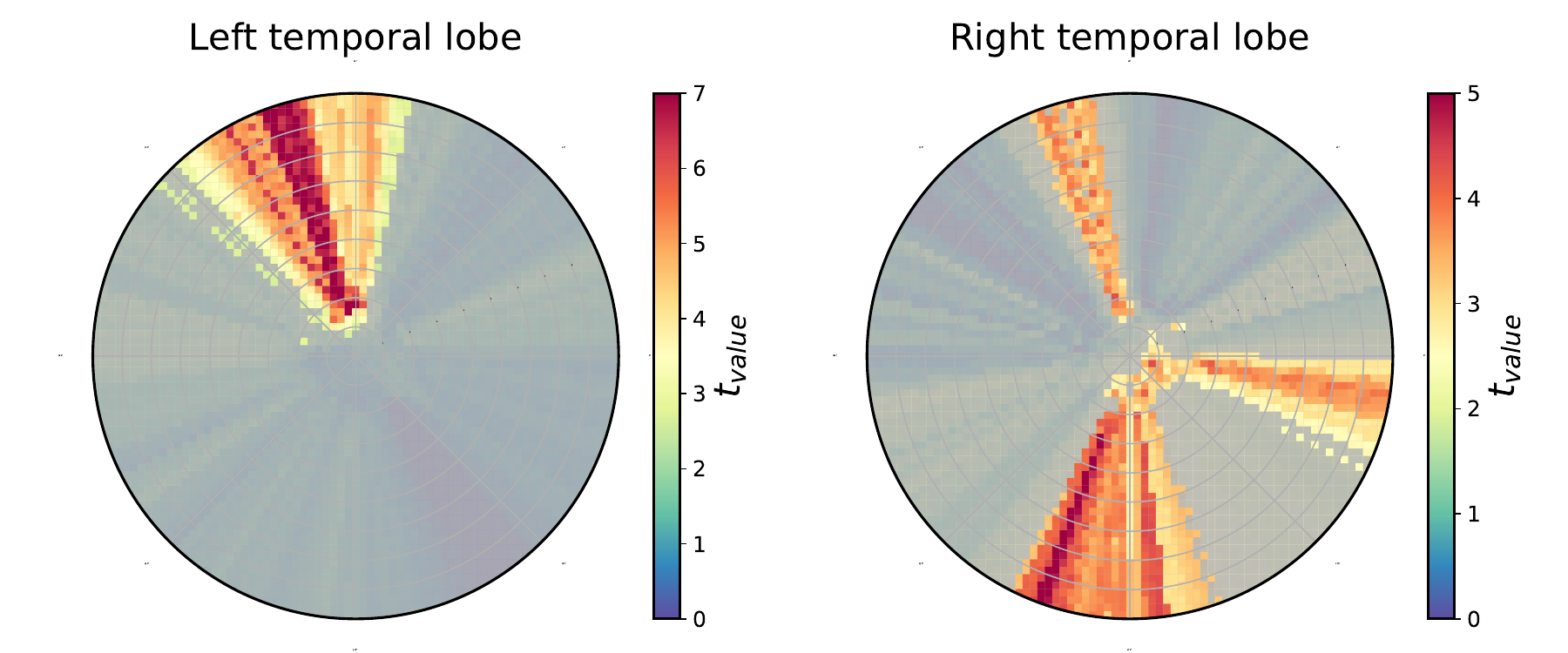}
    \caption{Student’s t-test comparison of interpolated HypDisp score values between patients with favorable and poor outcomes. Left disk: patients operated on the left temporal lobe; right disk: patients operated on the right temporal lobe. Non-shaded areas represent ROIs with significant differences between the two groups ($p<0.05$).}
    \label{HD_ROIS}
    \end{center}
    \end{figure}

The interpolated $\mathrm{HypDisp}$ scores from the two patient groups (good and poor outcomes) were compared inside the embedding disk using a Student’s t-test to identify regions in the disk (referred to as ``Region of Interest'' or ROI) with significant differences  ($p\le 0.05$). As shown in Fig.~\ref{HD_ROIS}, for patients with left hemisphere surgery, the ROI consists of a single component, while for those with right hemisphere surgery, three ROIs are identified. It is worthy of notice that the representation of networks in the Euclidean space (via diffusion maps~\cite{von2007tutorial}) did not reveal any discriminating node associated to the surgical outcome. These results suggest that embedding brain networks in the hyperbolic disk effectively capture surgery-related differences in brain connectivity associated with outcomes. \rev{For more clarity, the mean disk of each group are plotted Supp.~Fig.~2.}

As seen in Fig.~\ref{data_Embedding}, the interpolated $\mathrm{HypDisp}$ values clearly delineate a region in the hyperbolic disk where surgery mainly alters network structure. Notably, the surgical region does not entirely overlap with the regions whose connectivity is associated to the surgical outcome, indicating that surgery, as a perturbation of the connectivity graph, affects both local and global network structures.

\subsubsection{Balanced accuracy}
\rev{Given the class imbalance present in our dataset (i.e., unequal number of favorable and unfavorable outcomes), we report the \textit{balanced accuracy} in addition to the standard accuracy. Balanced accuracy is particularly suitable in this context, as it accounts for the performance on both classes equally, regardless of their proportions in the dataset.}

\rev{Formally, the balanced accuracy is defined as the average of the true positive rate (TPR) and the true negative rate (TNR):}

\begin{equation}
  \text{Balanced Accuracy} \;=\; \frac{1}{2}\!\left(
      \frac{\text{TP}}{\text{TP} + \text{FN}}
    + \frac{\text{TN}}{\text{TN} + \text{FP}}
  \right)
\end{equation}

\rev{where TP, TN, FP, and FN represent the number of true positives, true negatives, false positives, and false negatives, respectively. This metric ensures that both sensitivity (recall of the positive class) and specificity (recall of the negative class) are taken into account, which is essential in medical applications where both false positives and false negatives can have significant clinical implications.}

\subsection{Surgery outcome prediction}

To assess the impact of embedding alignment on surgery outcome prediction, we repeated the leave-one-patient-out procedure multiple times with different control network references. Evaluating the model across various reference networks from the healthy group, we obtained an Area Under the Curve (AUC) of  $0.87 \pm 0.003$ and a balanced accuracy of $0.81\pm 0.02$ (mean value $\pm$ SD). The reduced variability in performance indicates that the choice of reference network had no effect on outcome prediction.  

Six patients from the favorable outcome group (n = 42) were misclassified as having a poor outcome. Three of these patients relapsed at years 3, 4, and 5, while three others withdrew after three years. Conversely, two patients from the poor outcome group (n = 9) were misclassified as favorable outcomes, and both became seizure-free at years 3 and 4. One of these had a marginal probability of $0.51$ of being classified as favorable. Prediction analysis for each subgroup yielded an  AUC=$0.90\pm0.003$ and a balanced accuracy of $0.84\pm10^{-16}$ for the 30 patients who underwent left hemisphere surgery, and an AUC of $0.80\pm0.01$ and accuracy of $0.79\pm0.05$ for the 21 patients who underwent right hemisphere surgery.

To assess the impact of connections number on predicted outcomes, we applied our model to graphs obtained by randomly rewiring the original networks while preserving the degree distribution. For 25\% rewiring, the model yielded a mean AUC=$0.59$ and a balanced accuracy of $0.52$. As the number of rewired connections increased, the model could no longer distinguish between the two groups (no ROIs identified). These results clearly demonstrate that the predicted outcome is influenced by changes at the level of the entire network, rather than merely the number of resected connections.

\subsection{Brain regions associated to the surgery outcome}
\begin{figure}
    \begin{center}
    \includegraphics[width=1\textwidth]{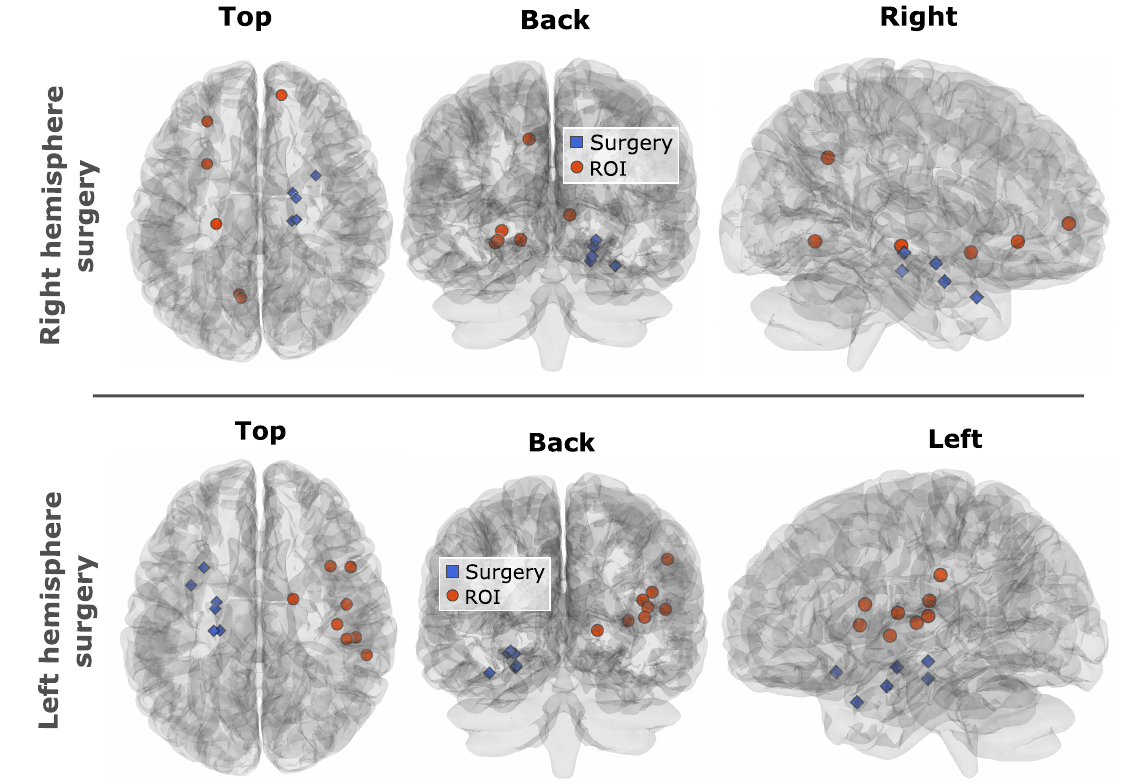}
    \caption{Projection onto brain space of the regions of interest from the hyperbolic disk showing the greatest outcome differences between groups. The first row shows patients operated on the right hemisphere, and the second row, those operated on the left. Blue points indicate brain areas directly affected by surgery, while red points represent key areas associated with outcome differences, as defined by the ROIs in the hyperbolic disk.}
    \label{brain_ROIS}
    \end{center}
    \end{figure}

To better interpret the nodes in the regions of interest, we back-projected them from the hyperbolic disk into brain space. Fig.~\ref{brain_ROIS} shows the nodes in brain space affected by the surgery and those in the discriminant ROIs of the hyperbolic disk. The nodes impacted by the surgery are concentrated in a small region of the operated hemisphere \rev{(In blue)}. For clarity, only nodes with $\geqslant 40$\% of connections removed are shown. In contrast, the nodes differentiating favorable and poor outcomes are mostly located in the contralateral hemisphere \rev{(In red)}, except for one node in the right surgery group. Visual inspection reveals clear hemispherical symmetry between regions impacted by the surgery. \rev{We define discriminant nodes as those located within the  region of interest (ROI) in at least 80\% of the patients. This criterion ensures that the selected nodes represent brain regions that are consistently affected by resection across the cohort.}

For patients who underwent left hemisphere surgery, the discriminant nodes associated to surgical outcome are concentrated in a small region of the contralateral hemisphere (Fig.~\ref{brain_ROIS}, bottom plots). In contrast, for those with right hemisphere surgery, the discriminant nodes are more dispersed in the left hemisphere (Fig.~\ref{brain_ROIS}, top plots). This contralateral localization in both groups highlights the importance of examining the large-scale network of each patient, beyond the local regions directly affected by surgery, to accurately assess its impact.

\section{Discussion}\label{Sec:discussion}
This study developed a framework to map and characterize the effects of surgery on neuroanatomical connectivity in epileptic patients. We explored embedding brain connectivity networks in non-Euclidean space to predict surgical outcomes in TLE patients. By mapping brain networks onto the hyperbolic disk, we identified connectivity changes that may serve as biomarkers for predicting surgical outcomes. Our results suggest that this framework can assess surgical resection impacts and predict outcomes by simulating disconnections in preoperative networks.

In patients with drug-resistant TLE epilepsy, current prognostic models combine clinical, neuroimaging, and electrophysiological data to predict surgical outcomes~\cite{Tonini2004, uijl2008prognosis}. However, no method has consistently demonstrated robust predictive power~\cite{uijl2008prognosis, Eriksson2023}. In this study, we explored embedding brain connectivity networks in hyperbolic space to predict surgical outcomes in TLE patients. Our findings show that mapping pre- and post-surgical networks in this non-Euclidean space reveals key connectivity discrepancies that could contribute to developing a predictive tool for surgical outcomes. Our approach, which analyzes brain networks in latent hyperbolic space, offers a novel framework for characterizing and mapping surgery's effects on neuroanatomical connectivity. We expect these low-dimensional, informative representations to be crucial for brain network studies, surpassing standard network measures and Euclidean embeddings. Notably, representations of pre- and post-surgery networks on Euclidean space fail to distinguish surgical outcome differences \rev{(See Supp.~Fig.~\textbf{1})}.

In a previous study, the use of pre-surgical fMRI connectivity networks provided an accuracy of  $76$\% for outcome prediction~\cite{he2017presurgical}. Similarly, preoperative brain connectivity from iEEG recordings predicted epilepsy surgery outcomes with $87$\% accuracy\cite{antony2013functional}. A connectivity-based simulation model achieved an AUC of $87$\%\cite{goodfellow2016estimation}, while a functional ECoG connectivity model predicted surgery outcomes with $83$\% accuracy~\cite{sinha2017predicting}. A dynamical iEEG model predicted surgical outcomes with an AUC of 89\% by extracting virtual resection network features~\cite{kini2019virtual}. Combining presurgical functional networks (iEEG) with structural connectivity (dMRI) yielded an AUC of $= 81$\% for predicting seizure outcomes~\cite{Sinha2023}. Using structural and functional MRI, connectivity anomalies predicted postsurgical seizure outcomes with 76\% accuracy~\cite{lariviere2020functional}. Evidence suggests that integrating neuroimaging data (including connectivity) with clinical information can improve predictions, achieving accuracy above 91\%~\cite{Memarian2015, Sinha2023}.

Prediction models based on anatomical connectivity from presurgical DTI data have yielded accuracy values ranging from 70\%\cite{munsell2015evaluation} to 83\%\cite{Bonilha2015}. However, these models focus on partial networks, specifically the ipsilateral temporal lobe's links to extratemporal regions~\cite{Bonilha2015}. A deep learning model using presurgical DTI-based connectomes achieved 88\% precision~\cite{Gleichgerrcht2018}. Using high-resolution reconstructions of cortical networks, a local connectivity group selection predicted surgical outcomes with 95\% accuracy, compared to 88\% for low-resolution parcellations~\cite{chen2021connectivity}. Alternatively, quantifying postsurgical  changes in dMRI connectivity predicted seizure outcomes with 79\% accuracy~\cite{taylor2018impact}. Another study used postsurgery connectivity changes to predict outcomes with an AUC of 84\%~\cite{Sinha2023}.  

Our prediction performances (AUC=$87$\% and balanced accuracy of $81$\%)  are comparable to these studies, though we note that some of the previously reported accuracy values may be biased due to an overrepresentation of patients with favorable outcomes. In contrast, our study addressed this imbalance by utilizing balanced accuracy, ensuring a more reliable assessment. Interestingly, the prediction model performed better for patients who underwent left hemisphere surgery (mean AUC=$0.90\pm0.003$), identifying a more compact region of interest in the brain. This aligns with studies showing differences between left and right hemisphere surgeries~\cite{Pustina2014, Fernandes2014}. These differences may be due to larger fiber tracks in left TLE patients compared to right TLE patients and healthy subjects~\cite{ahmadi2009side, besson2014structural}. We identified brain regions potentially involved in surgical failure, aiding in the differentiation of patient groups with distinct outcomes. Notably, the discriminant regions were consistently located in the contralateral hemisphere. While contralateral effects are not widely described, they are observed in TLE. Studies have shown contralateral abnormalities in preoperative DTI~\cite{Concha2005}, positron emission tomography (PET)~\cite{cahill2019metabolic}, and interictal scalp EEG data~\cite{Pustina2014}. These abnormalities have also been noted in baseline imaging and electrophysiological data~\cite{Jehi2010}. Brain connectivity studies have found contralateral differences linked to seizure recurrence, with TLE patients showing more connections in the contralateral hemisphere compared to healthy controls~\cite{he2017presurgical}, along with reduced long-range connections~\cite{lariviere2020functional, DeSalvo2020}.This reduction in contralateral centrality was also observed in DTI-derived structural networks~\cite{gleichgerrcht2020temporal}.  Additionally, metabolic network analysis using PET data has linked contralateral connectivity differences to surgical failure~\cite{stryvcek2024metabolic}. Our findings indicate that connectivity alterations associated with surgery exert a broad influence at the level of the entire network.

Our approach has some limitations that future studies should address. First, brain networks here are based on structural connections derived from MRI and dMRI data. However, structural imaging techniques generally assume bi-directional connections, which may inaccurately represent brain connectivity in primates~\cite{markov2014weighted}. Additionally, deterministic tractography methods can yield connectivity errors due to the crossing fiber problem~\cite{mori2002fiber}. Probabilistic tractography should be preferred for more reliable connectivity networks~\cite{bonilha2015reproducibility}. To estimate structural connectivity, MRI data is mapped to networks depicting anatomical connections between a reduced number of brain areas, typically based on an atlas. However, the method of node determination may affect network estimates of brain connectivity~\cite{zalesky2010whole}. Different brain atlases should be therefore evaluated to create networks with more nodes, enhancing spatial resolution~\cite{fornito2016fundamentals}, and predictive capabilities~\cite{chen2021connectivity}. Ultimately, Our methodology, based on anatomical brain networks in TLE, can be extended to connectivity networks from other imaging modalities (e.g., fMRI, iEEG) and extra-temporal epilepsies~\cite{reijneveld2007application, smith2011network, stam2014modern}. Furthermore, studying non-invasive connectivity networks (e.g., from EEG and MEG data) could provide a promising presurgical clinical tool.

Our results should be viewed in the context of a limited-size dataset from a previously published study~\cite{Sinha2021}. Postoperative network changes were able to statistically differentiate between favorable and poor outcomes only within the first year after surgery. However, for this limited dataset, creating more refined patient groups, such as those experiencing relapse at three or five years, would have led to more moderate differentiation and prediction. While short-term failure may result from incomplete resection, long-term relapse is influenced by many factors, including changes in medical treatment, lifestyle, and other variables not included in the current database~\cite{bell2017factors}. Incorporating connectivity markers from larger patient groups, along with clinical data, should improve long-term predictions~\cite{Eriksson2023}.

Computational models informed by brain imaging have provided insights into the networks involved in seizure generation and propagation~\cite{goodfellow2016estimation, sinha2017predicting, kini2019virtual}. Using patient-specific structural connectivity from DTI data, network-based simulations have predicted post-surgical outcomes through simulated resections~\cite{hutchings2015predicting}. Our approach offers an alternative to dynamic models for assessing the impact of disconnections and predicting seizure outcomes by performing in silico resections. If validated in larger studies, it could improve localization of epileptogenic networks, enhance surgical outcome predictions, and aid in estimating post-injury or post-intervention network changes, leading to better follow-up and prognosis.

\section{Methods}\label{Sec:methods}
\subsection{Dataset}
The dataset includes 51 patients who underwent anterior temporal lobe resection for epilepsy and 29 healthy subjects~\cite{Sinha2021}. Patients are divided into two groups: those who had left (n=30) or right (n=21) temporal lobe surgery. Patients were followed for five years post-surgery and classified according to the International League Against Epilepsy (ILAE) seizure outcome scale at annual intervals~\cite{Wieser2001}. Some patients in the favorable outcome group (seizure-free) experienced relapses at one, two, three, four, or five years~\cite{Sinha2021}. Due to the dataset's limited size, we focused on the ILAE outcome at one year, resulting in two groups: i) 9 patients (2 males, 7 females) with poor outcomes (ILAE~3-5) who continued to experience seizures, and ii) 34 patients (16 males, 18 females) with favorable outcomes (ILAE~1, seizure-free), plus 8 patients (2 males, 6 females) who had auras but no seizures (ILAE~2). Due to the small sample size, we did not predict outcomes for subgroups at two, three, or five years. For a complete description of the patients' demographic and clinical data, see Ref.~\cite{Sinha2021}.

\begin{table}[h!]
\centering
\begin{tabular}{lcc}
\toprule
\textbf{Variables} & \textbf{ILAE 1-2} & \textbf{ILAE 3-5} \\
\midrule
Patients, n & 42 & 9 \\
Sex (male/female), n & 18/24 & 2/7 \\
Age at onset (mean $\pm$ SD), y & 13.2 $\pm$ 10.3 & 19 $\pm$ 12 \\
Age at surgery (mean $\pm$ SD), y & 38.6 $\pm$ 11.9 & 46.5 $\pm$ 10.2 \\
Side (left/right), n & 22/20 & 5/4 \\
Hippocampal sclerosis, n (\%) & 24 (70.5) & 5 (55.5) \\
AEDs before surgery (mean $\pm$ SD), n & 6.3 $\pm$ 2.4 & 9.2 $\pm$ 3.3 \\
Preoperative MRI (normal/abnormal), n & 5/29 & 2/7 \\
History of status epilepticus, n (\%) & 5 (15.7) & 3 (33.3) \\
\bottomrule
\end{tabular}
\caption{Summary of demographic and clinical data of patients}
\label{tab:clinical_data}
\end{table}

\subsubsection{Brain networks}
The brain connectivity networks in this study were derived from anatomical neuroimaging data. Diffusion-weighted magnetic resonance imaging (dMRI) and structural MRI were performed pre-surgery, with only MRI conducted postoperatively. Pre-surgery networks were constructed using pre-surgery MRI and dMRI data. Post-surgery networks were generated by removing the resected brain regions from the pre-surgery dMRI using post-surgery MRI. The connections between brain regions were then identified through the dMRI data, revealing the underlying white matter fiber pathways. Data were discretized using a brain atlas of 114 regions (nodes). Further details on data acquisition and network reconstruction are given in Ref.~\cite{Sinha2021}.

\subsection{Network's embedding on the hyperbolic space}
\label{section_embedding}
Hyperbolic geometry provides a natural framework for embedding complex networks due to its ability to efficiently capture hierarchical and tree-like structures. In the hyperbolic disk model, nodes of a graph are mapped to points within a disk, where distances grow exponentially as they approach the boundary. The hyperbolic distance $\mathrm{dist}_{hyp}(i,j)$ between each pair of nodes $i$ and $j$, assigned with radii $(r_i, r_j)$ and angles $(\theta_i, \theta_j)$ at coordinates $(r_i,\theta_i)$ and $(r_j,\theta_j)$ in the disk, is computed according to the hyperbolic law of cosines \cite{Kitsak2020}:
\begin{equation}  
\label{eq_hyp_distance}
    \cosh \mathrm{dist}_{hyp}(i,j) = \cosh r_i \times \cosh r_j 
    - \sinh r_i \times \sinh r_j \times \cos(\pi-\vert \pi-\vert \theta_i-\theta_j\vert \vert )
\end{equation} 

Various techniques  (e.g., Mercator~\cite{GarciaMercator1019} , HyperMap~\cite{papadopoulos2014network}, or Hydra~\cite{keller2020hydra}, among others) have been developed for projecting graphs into hyperbolic space. These methods typically project the graph onto a hyperboloid, which is then mapped onto a 2D hyperbolic space model like the Poincaré or Klein disk. In this study, we project our networks directly onto the hyperbolic disk $\mathbb{D}^2$ using the coalescent embedding method~\cite{Muscoloni2017}, a machine learning-based approach known for its versatility and computational speed~\cite{Longhena2024}.  Starting with a binary connectivity graph, this method assigns effective edge weights using a repulsion-attraction rule that prioritizes edges with a significant role in information transmission~\cite{Muscoloni2017}: $\omega_{ij} = \frac{d_i+d_j+d_id_j}{1+CN_{ij}}$,  where $d_i$ is the degree of node $i$ and $CN_{ij}$ is the number of common neighbors between node $i$ and $j$.  The resulting network $\omega_{ij}$ is then projected onto the two-dimensional disk $\mathbb{D}^2$ using Isomap~\cite{von2007tutorial, Muscoloni2017}, a nonlinear dimensional reduction technique. The angular coordinates of the embedded nodes are adjusted uniformly, while maintaining their angular order. Finally, the radius of each node is determined by its rank in descending node degree: $r_i=\frac{2}{\zeta} (\beta \ln i + (1-\beta) \ln N)$, $i=1,2,...,N$;  where $N$ denotes the number of nodes, $\zeta$ a parameter determining space curvature, and $\beta$ a fading parameter. Here, we used $\zeta=1$ and $\beta=0.9$. Finally, we rescale the radial coordinates to fit the embedded nodes into a unitary disk. 

This method maps network's nodes to points within the disk $\mathbb{D}^2$, with radial coordinates reflecting the degree of centrality of each node. Nodes closer to the center are more central, while angular coordinates represent the degree of similarity between nodes. Nodes with smaller angular distances are more interconnected or similar. This approach combines radial centrality and angular similarity, providing a compact representation of both hierarchical and relational structures in the network~\cite{Muscoloni2017}.

Before comparing pre- and post-surgery brain connectivities, the embedded networks were realigned, as very often, a small perturbation may introduce a random angular offset to the nodes’ positions in the hyperbolic disk. A weak perturbation may, thus, result in two embeddings with similar connectivity between nodes but with a different structure regarding the embedding coordinates.  All networks were aligned with a reference connectivity network randomly drawn from the healthy control group. The alignment was here achieved by rotating the embedding data until the minimum sum of $\mathrm{HypDisp}$, calculated between all the nodes of the whole reference and aligned embedded networks, is reached. To assess the effect of the reference network, we performed the outcome prediction using various reference networks from the healthy group.

\subsection{Comparison of pre- and post-surgery networks}
To compare the brain networks before and after surgery, and to elucidate the local impact of the surgical procedure, we employ the $\mathrm{HypDisp}$ score, as described in~\cite{Longhena2024}. After aligning the two embeddings, the score assigned to each node is given by its hyperbolic displacement within the disc between the pre- and post-surgery networks. The score $\mathrm{HypDisp}(i)$ attributed to each node $i$ is given by :
\begin{equation}
    \mathrm{HypDisp}(i) = \mathrm{dist}_{hyp}\left(\mathrm{Pos_o(i)}, \mathrm{Pos_p(i)}\right)
\end{equation}
where $\mathrm{Pos_o(i)}$ and $\mathrm{Pos_p(i)}$ denote the position on $\mathbb{D}^2$ of node $i$ from the original and perturbed networks, respectively. $\mathrm{dist}_{hyp}\left(a,b\right)$ is the hyperbolic distance (Eq.~\ref{eq_hyp_distance}) between the two points $a$ and $b$ on the disk. \rev{The HypDisp score quantifies the local topological disruption experienced by each node following a network alteration, by measuring the dispersion of its embedding coordinates between the original and perturbed configurations, thus highlighting the brain regions most impacted by resection.}\\  

\subsubsection{Interpolation of $\mathrm{HypDisp}$ scores in $\mathbb{D}^2$}
\label{section_interpolation}
To compare local perturbations between patient groups, we interpolated the nodes’ $\mathrm{HypDisp}(i)$ scores across the entire disk.The space was discretized into pixels, with the number of pixels balancing computation time and node count. Using too few pixels led to information loss, while an excessive number increased computation time without enhancing precision. Here, we used a regular grid of 80 by 80 pixels for discretization, discarding pixels outside the disk. While alternative pixel distributions (e.g., hyperbolic) were possible, they had no significant impact on statistical comparisons.

To reduce the computational burden, the value assignated to each pixel $j$ was calculated as the weighted average of the $\mathrm{HypDisp}$ scores from its $k$ closest nodes (with the pre-surgery positions) on the disk $\mathbb{D}^2$:
\begin{eqnarray}
    C(j) = && \left(\sum _{i=1}^k  \frac{1}{{d_{\mathrm{hyp}}\left(\mathrm{Pos}(j), \mathrm{Pos_o}(i) \right)}^{\alpha}}\right)^{-1} \nonumber \\
    && \quad \times ~\sum _{i=1}^k \frac{1}{{d_{\mathrm{hyp}}\left(\mathrm{Pos}(j),\mathrm{Pos_o} (i) \right)}^{\alpha}} \mathrm{HypDisp}(i)
    \label{interpolation_formula}
\end{eqnarray}

Weights of each neighbor are given by the inverse of their hyperbolic distance to the pixel $j$ with the power $\alpha$. The more a node is distant from a pixel, the less its value contributes to the final averaged value. The two parameters $k$ and $\alpha$ do not depend on the data, and were set here to $k=20$ and $\alpha = 0.1$. One can notice that for low values of $\alpha$, the interpolation results in a very smoothed map. Conversely, large values of $\alpha$ assigns a greater influence to the closest neighbors of the interpolated pixel, resulting into a map formed by a mosaic of tiles.  Similarly, low values of $k$ yield irregularity in the interpolation, whereas an interpolation over a large number of neighbors is highly costly. Through experimentation, the value of $k=20$ was found to reduce the computational burden without significantly affecting the smoothness of the interpolation.  

\subsection{Statistical prediction of the surgery outcome}
Due to the limited database size, we used a leave-one-patient-out approach to assess our model's predictive capabilities. This method involves removing one patient, training the model on the remaining patients to identify ROIs in $\mathbb{D}^2$, and testing the model on the removed patient. \rev{For each patient, the median value of the interpolated HypDisp scores across all pixels within the previously defined group-level ROI on the hyperbolic disk was used as a feature in a logistic regression model to predict the probability of a favorable surgical outcome.} This process was repeated for each patient.

Prediction performance was evaluated using the area under the receiver operating characteristic curve (AUC) and balanced accuracy. Since accuracy can be biased toward the majority class (group of patients with favorable outcomes), we used balanced accuracy, which combines sensitivity and specificity, to account for the dataset imbalance.

\bmhead{Acknowledgements}
M.G. and A.L acknowledge financial support from the Ecole Normale Supérieure de Lyon and the doctoral school EDITE from Paris Sorbonne University, respectively.

\bmhead{Conflict of interest}
V. Navarro reports fees from Boards with UCB Pharma, EISAI, Liva Nova, GW Pharma. The remaining authors have no conflicts of interest. 

\bmhead{Data availability}
The brain networks used in this study are publicly available in the supplementary material of Ref.~\cite{Sinha2021}.

\bmhead{Authors contribution}
M.G. and M.C.: conceptualization, methodology, investigation, result visualization, and writing-original draft; A.L., L.C. and V.N..: methodology, writing-review and editing; V.F. and B.M.: conceptualization, writing-review and editing

\bibliography{biblio}

\includepdf[pages=-, fitpaper=true]{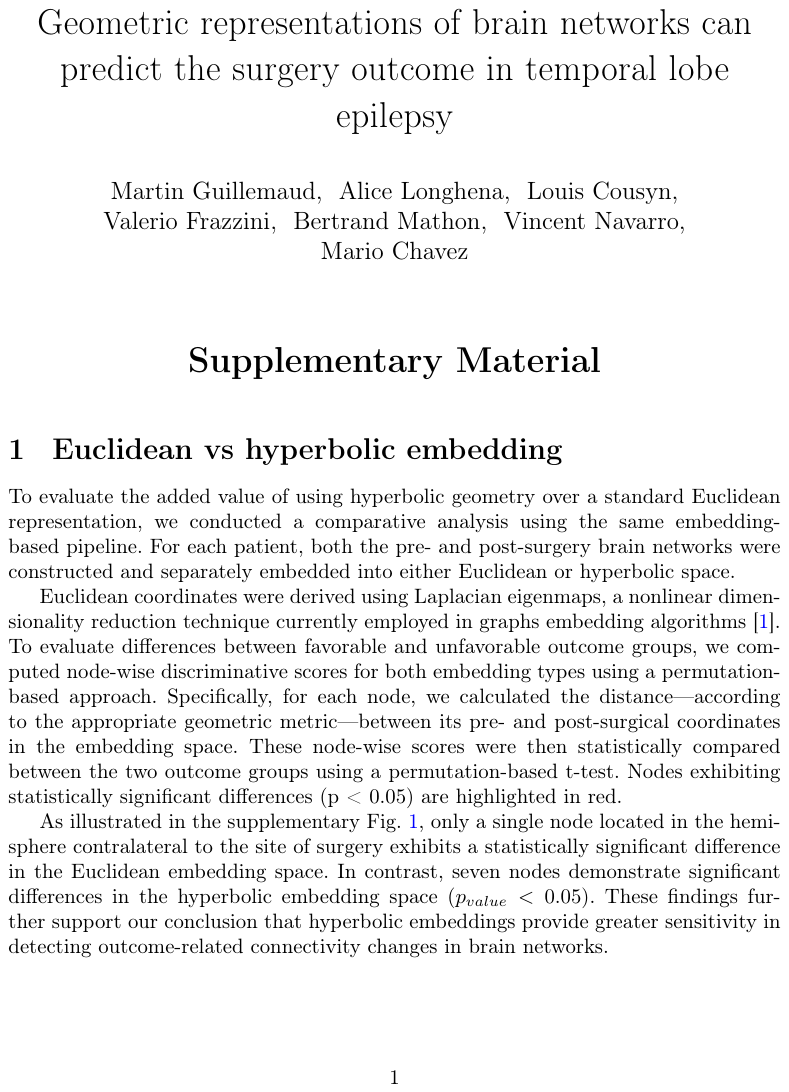}

\end{document}